\def\ket{\vert \vert	\{ \emptyset \} \rangle}
  \def\ket2{\vert \vert \otimes \{ R \} \rangle}

\def\.#1{\mathaccent 95#1}
\def\^#1{\mathaccent 94 #1}
\def\~#1{\mathaccent "7E #1}

\def\eq{\enskip =\enskip}
\def\pls{\enskip +\enskip}
\def\mns{\enskip -\enskip}

\def\trans{{\cal T}}
\def\proj{{\cal P}}

  \def\proj{{\cal P}}
  \def\trans{{\cal T}}
  \def\ket{\vert \vert	\{ \emptyset \} \rangle}
  \def\ket2{\vert \vert \otimes \{ R \} \rangle}

\def\.#1{\mathaccent 95#1}
\def\^#1{\mathaccent 94 #1}
\def\~#1{\mathaccent "7E #1}

\def\eq{\enskip =\enskip}
\def\pls{\enskip +\enskip}
\def\mns{\enskip -\enskip}

\documentstyle[12pt]{ioplppt}
\begin{document}
\title{ Electronic structure and magnetism 
of disordered bcc Fe alloys}
\author{\bf Subhradip Ghosh, Biplab Sanyal$^{\dagger}$, Chhanda Basu Chaudhuri 
 {\rm and} Abhijit Mookerjee}
\address{S.N.Bose National Centre for Basic Sciences,  JD Block, 
Sector 3,  Salt Lake City,  Calcutta 700091,  India\\
$^{\dagger}$ Department of Physics,  Brock University,  St. Catharines,  Ontario L2S 3A1,  Canada.}

\begin{abstract}
We here study electronic structure and magnetic properties 
of disordered bcc Co$_{x}$Fe$_{1-x}$, Cr$_{x}$Fe$_{1-x}$ and Mn$_{x}$Fe$_{1-x}$ alloys in their
ferromagnetic phases using
Augmented Space Recursion (ASR) technique coupled with tight-binding
linearized muffin tin orbital   (TB-LMTO) method.  
We calculate the densities of states, magnetic moments and Curie
temperatures of these alloys to show the
variation upon alloying Fe with the other neighbouring 3-{\it d} 
transition metals 
using arguments based on charge transfer, exchange splitting and
hybridization effects.
\end{abstract}
{\parindent 0pt
\baselineskip 25pt

\section{Introduction}

 Properties of magnetic materials have been a subject of great 
 scientific and practical interest. An enormous amount of experimental
 and theoretical investigations has been carried out to have a proper
 understanding of the nature of magnetism in solids \cite{kn:wohl}.
 Fe, being one of the ferromagnets among the late transition metals has
 drawn considerable attention due to its interesting magnetic properties.
 Apart from studies on elemental Fe which include the studies on structural 
 \cite{kn:ishi}and magnetic phase stability \cite{kn:and, kn:pinski, kn:hirai}
 there are numerous investigations on magnetism in Fe based 
 ordered and disordered
 alloys, both theoretically and experimentally. The experimental
 investigations have provided a variety of information about the magnetic
 properties of these systems e.g. variation of magnetization with band 
 filling \cite{kn:shull, kn:bardos}, moment distribution in dilute Fe alloys
 in low \cite{kn:collins} as well as in finite temperatures \cite{kn:child}, 
 local environmental effects on magnetic properties \cite{kn:radha, kn:kaj}, 
 spatial distribution and thermal variation of hyperfine fields \cite{kn:arp}, 
 concentration dependence of high field susceptibility \cite{kn:stoel}, low
 temperature specific heat \cite{kn:cheng, kn:schroder} and magnetic phase
 stability leading eventually to magnetic phase diagrams \cite{kn:baal}.

 The earlier theoretical studies were based on various
 models of band structure calculations \cite{kn:has, kn:jo}. Though these
 calculations were successful too a certain extent in 
 explaining the experimental
 observations, they suffered from the drawback of having too many adjustable
 parameters which limited the reliability of their results. But, with the recent
 progresses in first principles electronic structure techniques, 
 the properties of magnetic alloys are investigated more accurately 
 and efficiently.
 Different aspects of magnetism in both ordered and disordered phases 
 of Fe alloys have been studied successfully by these techniques overcoming the
 limitations of earlier model calculations \cite{kn:james, 
 kn:burke, kn:akai, kn:pad}.
  In this communication, we aim at a systematic study of electronic structure
 and magnetic properties of substitutionally disordered Co$_{x}$Fe$_{1-x}$, 
 Cr$_{x}$Fe$_{1-x}$ and Mn$_{x}$Fe$_{1-x}$ alloys using the self-consistent
 TBLMTO-ASR technique. In the elemental phase, bcc Fe is a ferromagnet, 
 bcc Cr is a weak non-commensurate antiferromagnet, Mn has a very
 complicated crystal   (unit cell of 58 atoms) and magnetic structure while Co
 is a ferromagnet.  
 For FeCr, Fe atoms stabilize the
 commensurate antiferromagnetic   (B2) order in the  Cr-rich side   (x$>$0.8) although
 Cr$_{x}$Fe$_{1-x}$ with x$>$0.8 are ferromagnets and stabilize in bcc lattice
 \cite{kn:kuli}. In case of MnFe alloys, ferromagnetic phase is stable only
 up to x=0.2 and the crystal stabilizes in bcc lattice. For x$>$0.2, several 
 phases with antiferromagnetic ordering get stabilized \cite{kn:endoh}. In
 FeCo alloys, however,  the ferromagnetic phase is stable for the full range of
 concentrations and the crystal too stabilizes in the bcc structure. 
 To our knowledge, no systematic 
 study  on these three systems has been done so far using the same methodology, 
though they are
 four successive members on the same row of periodic table. This motivated
 us to perform a systematic investigation of these systems. 
  In this work, 
 we have restricted ourselves only to the ferromagnetic phases of these systems.

\section{Theoretical Details}
To study these alloys we will use the methodology of the 
augmented space recursion technique \cite{kn:as, kn:saha, kn:bs}
in the first principles framework of tight-binding linearized muffin 
tin orbital method \cite{kn:oka}. Extensive details of the
description of the effective augmented-space Hamiltonian have been given
in an earlier paper\cite{kn:ppb}. Here we shall quote the key results of
generalized TBLMTO-ASR . 

\begin{eqnarray}
H \eq \sum_{RL} \^C_{RL} \proj_{RL} \pls \sum_{RL}\sum_{R'L'} \^\Delta^{1/2}_{RL}
S^{\beta}_{RL, R'L'}\^\Delta^{1/2}_{R'L'} \trans_{RL, R'L'} \nonumber \\
\^C_{RL} \eq C^{B}_{RL} \pls \left  (C^{A}_{RL}-C^{B}_{RL}\right)\;n_{R} \nonumber\\
\^\Delta^{1/2}_{RL} \eq \Delta^{B1/2}_{RL}\pls \left  (\Delta^{A1/2}_{RL} - \Delta^{B1/2}_{RL}
\right)\; n_{R}
\end{eqnarray}

Here $\proj_{RL}$ and $\trans_{RL, R'L'}$ are projection 
and transfer operators in the Hilbert
space spanned by the tight binding basis $\vert RL\rangle$ 
and $n_{R}$ is a random occupation
variable which is 1 if the site $R$ is occupied by an atom 
of the A type and 0 if not. 
$C^{Q}_{RL}$ and $\Delta^{Q}_{RL}$ are potential 
parameters describing the scattering properties of the
constituents(Q=A, B) of the alloy and $S^{\beta}_{RL, R'L'}$ 
is the screened structure constant describing the geometry of the
underlying lattice. The augmented
space Hamiltonian replaces the random occupation 
variable by operators $M_{R}$ of rank 2. For
models without any short-range order

\[ M_{R} \eq x \proj_{\uparrow}^{R} \pls   
(1-x) \proj_{\downarrow}^{R} \pls \sqrt{x  (1-x)}
\left  ( \trans_{\uparrow\downarrow}^{R}+\trans_{\downarrow\uparrow}^{R} \right) \]

\[ \vert \uparrow \rangle \eq \left  ( \sqrt{x}\vert 0 \rangle \pls \sqrt{1-x}\vert 1 \rangle \right) \]
\[ \vert \downarrow \rangle \eq \left  (\sqrt{1-x}\vert 0\rangle \mns \sqrt{x}\vert 1 \rangle \right) \]

The  recursion method then expresses the Green functions as continued fraction expansions. The continued
fraction coefficients are exactly obtained up to eight levels and the terminator suggested by Luchini
and Nex \cite{kn:ln} is used to approximate the asymptotic part. The convergence of this procedure has
been discussed by Ghosh \etal \cite{kn:gdm}. The local charge densities are given by :

\begin{equation}
\rho^{\lambda}_{\sigma}  (r) \eq   (-1/\pi) \Im m \sum_{L} \int_{-\infty}^{E_{F}} dE \ll G_{LL}^{\lambda, \sigma}  (r, r, E)\gg
\end{equation}

Here $\lambda$ is either $A$ or $B$. The local magnetic moment is

\[ m^{\lambda} \eq \int_{r<R_{WS}} d^{3}r\; \left[\rho_{\uparrow}  (r)\mns \rho_{\downarrow}  (r)\right] \]

The Curie temperature $T_{C}$ can be calculated using Mohn-Wolfarth model \cite{kn:mw} from the expression 
$$  \frac{T_{C}^2}{T_{C}^{S^2}}+\frac{T_{C}}{T_{SF}} \mns 1 \eq 0$$
where, 
$T_{c}^S$ is the Stoner Curie temperature calculated from the relation
\begin{equation}
\langle I  (E_{F}) \rangle \int_{-\infty}^{\infty} N  (E) 
\left  (\frac{\delta f}{\delta E} \right) dE \eq 1
\end{equation}
$\langle $I(E$_{F}$)$\rangle$ is the concentration averaged Stoner parameter. The parameters of pure elements are 
 obtained from the earlier calculations
\cite{kn:jan}
,  $N  (E)$ is the density of states per atom per spin \cite{kn:gun}
and $f$ is the Fermi distribution function.
$T_{SF}$ is the spin fluctuation temperature given by, 
\begin{equation}
T_{SF} \eq \frac{m^2}{10k_{B} \langle \chi_{0}\rangle} 
\end{equation}
$ \langle \chi_{0} \rangle$ is the concentration weighted exchange enhanced spin susceptibility at equilibrium and $m$ is
the averaged magnetic moment per atom.
$\chi_{0}$   (pure elements) is calculated using the relation by Mohn \cite{kn:mw} and Gersdorf
\cite{kn:ger}:

\[\chi_{0}^{-1} \eq  \frac{1}{2\mu_{B}^2}\left  (\frac{1}{2N^\uparrow  (E_{F})}\pls
\frac{1}{2N^\downarrow  (E_{F})} \mns I\right)\]

$I$ is the Stoner parameter for pure 
elements and $N^\uparrow  (E_{F})$ and $N^\downarrow  (E_{F})$ are
the spin-up and spin-down partial density of states 
per atom at the Fermi level for each species in the alloy.

\section{Calculational Details}

For our calculations 
 we have used a real space cluster of 400 atoms and
an augmented space shell up to the sixth nearest neighbour from the starting
state. Eight pairs of recursion coefficients were determined exactly and the
continued fraction terminated by the analytic terminator due to Luchini
and Nex \cite{kn:ln}. In a paper Ghosh \etal \cite{kn:gdm} have shown the
convergence of the related integrated quantities,    like the Fermi energy,  
the band energy,   the magnetic moments and the charge densities,   
within the augmented space
recursion. The convergence tests suggested by the authors were carried
out to prescribed accuracies. We noted that at least eight pairs
of recursion coefficients were necessary to provide Fermi energies
and magnetic moments to required accuracies. We have reduced the 
computational burden of the
energy dependent recursion method 
using the seed
recursion methodology \cite{kn:gm} with fifteen energy seed points
uniformly across the spectrum.

We have varied the Wigner-Seitz radii of the two constituent atoms 
in order to have charge neutral spheres. This eliminates the
necessity to calculate the Madelung energy 
which is a difficult task for the case of disordered alloys.
 Simultaneously we have made sure that the sphere overlap remains
within the 15$\%$ limit prescribed by Andersen.

The calculations have been made self-consistent in the LSDA sense,   that
is,   at each stage the averaged charge densities are calculated from the
augmented space recursion and the new potential is generated by the
usual LSDA techniques. This self-consistency cycle was converged in
both total energy and charge to errors of the order 10$^{-5}$. 
The exchange-correlation potential of Von Barth and Hedin has been used, 
$s, p$ and $d$ orbitals were used to construct the basis functions and scalar
relativistic corrections were included. For all the calculations,
we have used the lattice constants for the alloys according to
Vegard's law.

\section{Results and Discussion}
In our approach we emphasize on interrelations of magnetism and charge
transfer behaviour. 
Since in our calculations
we have maintained local charge neutrality, we have to deal with the question
of strong variation of magnetic moments. Within the itinerant electron theory
of magnetism this can be understood in terms of a redistribution of local
electronic charge either between two spin directions. Together with Coulomb
interaction which determines the positions of atomic {\it d} levels of the
constituents and thus the charge transfer in case of a transition metal alloy, 
magnetic exchange and hybridization play very important role in determining
the magnetic properties. This has already been observed in certain cases
\cite{kn:rich, kn:schwarz}. 

These facts can be expressed in a more quantitative form using {\it d}-orbital
potential parameters C$_{d\sigma}^{Q}$ obtained from TBLMTO for both alloy
components  (Q=A, B) and for both spin directions  ($\sigma$=$\uparrow$, $\downarrow$).
These quantities are equivalent to the atomic {\it d} levels.

The spin dependent diagonal disorder in a random binary alloy A$_{x}$B$_{1-x}$
can be defined as \cite{kn:kur} , 

\begin{equation}
$$\delta_{\sigma}=C_{d\sigma}^{A}-C_{d\sigma}^{B}$$
\end{equation}

The local exchange splitting can be defined as \cite{kn:kur}, 

\begin{equation}
$$\Delta_{e}^{Q}=C_{d\downarrow}^{Q}-C_{d\uparrow}^{Q}$$
\end{equation}

Figure 1 shows the compositional dependence of local and average magnetic
moments for Co$_{x}$Fe$_{1-x}$ alloys. The filled 
triangles denote the average moments
while the filled circles denote Fe moment and filled squares denote Co moments.
The open triangles, 
 open circles and the open squares denote experimental 
values of the averaged moments \cite{kn:bardos}, Fe local
moments \cite{kn:coll} and the
Co local moments \cite{kn:coll}respectively. It is clear that
our results agree well with the experiments, in particular the qualitative 
trend of local as well as average magnetic moments. The results show that
the Fe local moment increases with increasing Co content up to x=0.3 beyond
which it tends towards a saturation while the Co moment remains almost
constant for the whole concentration regime. As a result, the average 
magnetization reaches a maximum at 30$\%$ of Co beyond which it starts 
to decrease. Similar behaviour has been observed in previous studies using
LCAO-CPA \cite{kn:rich} and LMTO-CPA \cite{kn:turek}.
This non-monotonic variation of average magnetization can be explained from
the variation of local number of electrons  (Figure 2) and density of states
  (Figure 3). In this case, a transition from weak  ferromagnetism  (incompletely
filled majority {\it d} band) for Fe-rich side to strong ferromagnetism
  (majority {\it d} band completely filled) for alloy with a higher Co content
  (x$>$0.3) is seen. The initial increase of alloy magnetization corresponds
to a continuous filling of majority bands while the minority bands remain
almost constantly occupied  (Figure 2  (a)). The linear decrease of alloy 
magnetization with increasing x for x$>$0.3 reflects a strong ferromagnetic
region in which majority bands are fully occupied whereas the minority bands
accommodate more electrons with increasing Co content.The filling of majority
band up to x=0.3 mainly occurs due to incompletely filled majority {\it d}
band of weak ferromagnet Fe  (Figure 2  (b)) while the rise in the minority
band filling beyond x=0.3 is essentially due to a fall in Fe minority
electrons and an almost constant nature of Co minority electron number
variation (Figure 2  (c)). This is reflected in features of the density of states
  (Figure 3) as well as variation of the density of states at Fermi level n (E$_{F}$) 
  (Figure 4  (a)). For majority spin states the filling of {\it d} band 
occurring at Fe sites is accompanied by a steady decrease of n (E$_{F}$). 
 Upto x=0.3, the Fermi level is pinned to the
minimum of minority spin density of states  (Figure 3  (a)-  (b)). Increasing Co
content, which essentially means gradual filling, shifts the Fermi level to regions
of low spin up density of states and above x=0.3 to increasing spin down
density of states  (Figure 3  (c)-3  (f)). Thus n($E_{F}$) for up spin   (shown by
up triangles in Figure 4  (a)) decreases while n($E_{F}$) for down spin   (shown
by down triangles) increases continuously beyond x=0.3. As a result, average
n($E_{F}$)   (shown by squares) goes through a minimum around x=0.3.

All these phenomena are a consequence of local charge neutrality, exchange 
 and hybridization. Since bcc 
Co is already a saturated ferromagnet, there is hardly any possibility to
increase substantially the number of majority spin electrons and thereby the
local magnetic moment of Co. Because of small $sp$-density of states at Fermi
level  (Figure 3) compared with the {\it d} contribution, the transfer of minority
spin {\it d} electrons to $sp$-states is expected to be very small. Thus, Co
moment is almost independent of alloy concentration and the possible exchange
splitting of Co {\it d} level remains almost constant throughout the concentration 
regime  (Table-1). On the other hand, the weak ferromagnetism of
 Fe gives rise to the possibility of filling approximately 0.3 majority spin holes
with minority spin electrons. Thus, local Fe moment increases as a result of
increase in local exchange splitting  (Table-1). Hence, inspite of being
nearest neighbours in the periodic table the exchange makes their behaviour
so very different.

\begin{table}
\centering
\begin{tabular} {ccc}
\hline
$x_{Co}$ &$\Delta_{e}^{Fe}$ &$\Delta_{e}^{Co}$\\
\hline
0.1 &0.179 &0.136\\
0.2 &0.185 &0.139\\
0.3 &0.190 &0.140\\
0.4 &0.194 &0.142\\
0.5 &0.196 &0.142\\
0.6 &0.199 &0.142\\
0.8 &0.201 &0.138\\
\hline\end{tabular}
\caption{ Local exchange splitting values  (in Ryd) in
Co$_{x}$Fe$_{1-x}$ alloys with varying
concentration of Co}
\end{table}

The role of hybridization influencing the local magnetic properties can be
explained in terms of the bonding charge transfer  (BCT) model \cite{kn:rich1}. As is evident from the
density of states, the disorder in the minority spin band is more prominent
which is also realized quantitatively from $\delta^{\sigma}$ variation.
While $\delta^{\uparrow}$ varies from 40 mRyd from Fe-rich side to 6 mRyd in
Co-rich side, $\delta^{\downarrow}$ remains $\sim$0.8 Ryd for the whole range of
concentrations. According to the BCT model, different positions of atomic 
$d^{\downarrow}$-levels of Fe and Co cause bonding charge transfer  (BCT) in
the minority spin band. An inspection of density of states at various
concentrations  (Figure 3) shows that the bonding part of spin-up density of
states has a larger Co weight whereas Fe dominates the anti-bonding part.
A transition of minority spin electrons from Fe to Co occurs. To retain local
charge neutrality, mainly Co minority spin electrons are transferred to Fe
majority band causing an increase of exchange splitting and magnetic 
saturation. As a result, a net electron redistribution from Fe$^{\uparrow}$
to Fe$^{\downarrow}$ state occurs only to increase Fe moment.

So, to conclude, the magnetization behavior of CoFe is characterized on the
Fe-rich side by the magnetic saturation due to hybridization whereas the Co-
rich side is determined simply by filling of minority band.

Figure 4  (b), 4  (c) and 4  (d) respectively show the variation of inverse 
susceptibility, spin fluctuation temperature and Curie temperature calculated
using MW model. The variation of inverse susceptibility is exactly reverse
in nature to that of n($E_{F}$). This is due to the fact that inverse
susceptibility is dependent on n($E_{F}$)$^{-1}$ only as stoner parameter is
a constant quantity. The variation of spin fluctuation temperature follows
the same nature as of magnetization and inverse susceptibility which are alike
and this nature is reflected in Curie temperature because though in the calculations
  of Curie temperature Stoner Curie temperature was also involved. The
diamonds in Figure 4  (d) represent the results of Stoner Curie temperature
  ($T^{S}_{C}$). Apart from much larger values the nature of variation is also
non-linear. This is due to the fact that $T^{S}_{C}$ measures the temperature
at which the paramagnetic state becomes unstable rather than the magnetic
transition temperature.

Figure 5 shows the variation of average and local magnetization in 
Mn$_{x}$Fe$_{1-x}$ alloys with Mn concentration.The filled triangles 
stand for average value while the filled circles and filled squares
represent Fe and Mn local moments respectively.
Unfortunately, enough
experimental data is not available in this region to support our results.
The experimental results so far available  (shown by open triangles) 
\cite{kn:fish} agree well with our results. Our results also agree to 
a reasonable extent with the calculations based on Hartree-Fock-CPA
\cite{kn:kan1} qualitatively but the variation of Mn local moment 
doesn't agree qualitatively with KKR-CPA results \cite{kn:kuli1}. In our case
the Mn local moment linearly decreases with increasing Mn concentration, 
a feature obtained in Hartree-Fock-CPA calculations too but the KKR-CPA
results predict opposite trend for Mn moment. The Fe moment weakly increases
and the average moment decreases which is in qualitative agreement
with the experimental Slater-Pauling curve \cite{kn:fish}.

These variations can be explained once again using band filling   (Figure 6)
and density of states  (Figure 7) results. Figure 6  (b) shows an almost
constantly filled Fe up and down bands across the concentration regime
thereby supporting the weak variation in Fe local moment. In case of
Mn,  the filling accommodates more number of electrons in the minority band
  (Figure 6  (c)). As a result,  the minority band of the alloy gets gradually
filled up while a loss of electrons from majority bands occur  (Figure 6  (a)).
This feature is manifested in density of states as well as in variation
of n($E_{F}$)  (Figure 8  (a)).For majority spin states the filling of
majority band at Fe site reduces n($E_{F}$) for the corresponding band
while increasing Mn content shifts Fermi level to regions of low spin
up density of states and high spin down density of states because of gradual
filling of minority electrons.

Once again, these phenomena can be explained on the basis of interplay of
local charge neutrality, hybridization and magnetic exchange. Unlike Co, 
Mn {\it d} level exchange splitting varies quite considerably  (variation of the order
of 30 mRyd) due to gradual filling of the minority band and de-filling of the majority
band. As a result, Mn local moment decreases as one goes to Mn rich region. 
In Fe,  since both the bands are nearly filled and a very weak variation
of number of local electrons is observed. The  local exchange splitting variation is of 
 the order of 14 mRyd only  (Table-2). Hence Fe local moment increases, 
though quite weakly compared to Mn.

\begin{table}
\centering
\begin{tabular} {ccc}
\hline
$x_{Mn}$ &$\Delta_{e}^{Fe}$ &$\Delta_{e}^{Mn}$\\
\hline
.05 &0.177 &-0.170\\
0.1 &0.183 &-0.187\\
.15 &0.188 &-0.197\\
0.2 &0.191 &-0.202\\
\hline\end{tabular}
\caption{ Local exchange splitting values (in Ryd) in
Mn$_{x}$Fe$_{1-x}$ alloys with varying
concentration of Mn}
\end{table}

The role of hybridization and charge re-distribution can be addressed as
follows. A look at density of states reveals that unlike FeCo, the disorder
is appreciable in both the bands. For the majority band,  $\delta$ increases
up to 50 mRyd while $\delta^{\downarrow}$ increases around 8 mRyd only. As
Mn concentration is increased
the bonding part of spin-up density of states is dominated by Fe while
the anti-bonding part is dominated by Mn. The reverse situation is observed
for spin down density of states. As a result,  majority spins from Mn migrate
to Fe and minority spins from Fe migrate to Mn. Thus an 
increase in Fe local moment is observed. Finally a transition of
electrons from Mn$^{\uparrow}$ to Mn$^{\downarrow}$ state occurs reducing
the Mn moment gradually.

Figure 8  (b),   (c) and   (d) respectively show
the results on inverse susceptibility,
spin fluctuation temperature and the Curie temperature $T_{C}$.
The inverse susceptibility shows
a non-linear behaviour and as a result unlike FeCo, we observe a variation
of spin fluctuation temperature exactly opposite to that of magnetic moment.
But, once again like FeCo,  $T_{C}$
reflects the behaviour of spin fluctuation
temperature though $T_{C}^{S}$
(shown by filled circles in Figure 8  (d))
behaves in a opposite way. These discrepancies are due to limitations of
$T_{C}^{S}$ itself which has been discussed for the case of FeCo alloys.

Figure 9 shows the concentration dependence of local and average magnetic
moments in Cr$_{x}$Fe$_{1-x}$ alloy. The solid up triangles represent
the calculated average values while the solid circles and solid squares
denote the Fe and Cr local moments respectively. Our average magnetization
results agree well with the experimental values  (shown by open up triangles)
\cite{kn:ling} and other theoretical results \cite{kn:kuli1, kn:but, kn:kan1}.
In fact, for higher Cr concentrations the experimental points almost fall
on the theoretical curve establishing good agreement. In case of local 
moments, our results for Fe agree considerably well with available experimental
\cite{kn:but} results but there is quantitative difference in Cr moment values
with those of earlier calculations \cite{kn:kuli1, kn:ded}. In our case, we
obtain a larger negative value of Cr moment which though increases rapidly
in the Cr-rich region but never changes its sign which has been observed
in earlier theoretical calculations around $x$=0.7. However, this slight discrepancy doesn't affect the average properties at all as is seen from the 
quantitative agreement with the experiments. Even the qualitative nature
of variation of local as well as average moments is well reproduced. As is
seen from the figure,  Fe moment remains almost constant up to
around $x$=0.4 and
then it decreases in the Cr-rich side but the nature of variation is pretty
weak. The Cr moment on the other hand increases rapidly as Cr content is
increased making the average value to drop down very fast and approaching
zero in accordance with the established observation that the average moment
collapses around $x$=0.8 due to
transition from ferromagnetic to antiferromagnetic state.

Once again, we take recourse to the density of states  (Figure 11) and 
variation of local and average number of electrons as number of valence
electrons is decreased  (Fe-rich to Cr-rich side)
(Figure 10) to explain these
behaviours. A thorough inspection of density of states for various 
concentrations show that $E_{F}$ is positioned in a valley between the
bonding and antibonding peaks in the minority spin density of states.
This feature explains
the reason for linear variation of average moment because as we keep on
increasing Fe content electrons are added to the majority spin states without
much affecting minority spins. This feature is very clear in Figure 10  (a).
The weak variation of Fe magnetic moment can also be explained likewise.
The partial Fe density of states for both spins show little variation
across the whole range of concentrations whereas the Cr minority density of 
states vary appreciably as we scan through the concentration regime. This is
understandable if one looks into the variation of majority and minority 
electrons at each site. In case of Fe  (Figure 10  (b)),
both the majority and
minority bands are almost completely filled while in
case of Cr  (Figure 10  (c)),
the majority band accommodates more and more electrons as Cr-content is
increased while the minority band loses and eventually they vary in such a
way that at a certain critical concentration there will be more number of
electrons in the majority band. Due to this behaviour of Cr, the average
number of up electrons decrease  (Figure 10  (a)) rapidly in
contrast to almost
constantly filled minority band in such a way that around $x$=0.8,
the number
of electrons in the majority band will be same as that of minority one
establishing a collapse of magnetic moment when ferromagnetism to
antiferromagnetism transition
will take place. n($E_{F}$) variation supports this
(Figure 12  (a)). Initially
the Fermi level is situated near the {\it d}-level peak in the majority
band  but as the Cr content is increased and majority band starts loosing
electrons Fermi level starts moving away from high density of states though
initially in the Fe-rich region due to increase in
Fe majority band electrons
up to $x$=0.4 n($E_{F}$) had a weak rise. But as we step into Cr-rich region
this effect is completely washed out. On the other hand since minority
band is almost filled there is almost no
variation in n($E_{F}$)$^{\downarrow}$.
As a result,  the average n($E_{F}$) has a maxima around $x$=0.4.

We now look for investigating the role of hybridization, exchange etc and
the type of charge distribution within the constraint of local charge
neutrality. In Cr, the local exchange splitting varies more strongly
than Mn and also in a opposite way. In case of Mn, exchange splitting 
decreased towards Mn-rich region whereas in this case, it increases as Cr
content is increased  (Table-3). The variation in the local exchange splitting
for Cr varies of the order of 64 mRyd from a Fe-rich to a Cr-rich region.
This is due to the rapid de-filling of Cr minority band. In Fe, since
both the bands are nearly filled,  the local exchange splitting does not
vary as much like that of Cr. Nevertheless, unlike Mn, it decreases and the
variation is of the order of 28 mRyd, explaining the decrease of Fe moment.

\begin{table}
\centering
\begin{tabular} {ccc}
\hline
$x_{Cr}$ &$\Delta_{e}^{Fe}$ &$\Delta_{e}^{Cr}$\\
\hline
.25 &0.182 &-0.080\\
0.4 &0.179 &-0.059\\
0.5 &0.176 &-0.047\\
0.6 &0.171 &-0.035\\
0.7 &0.163 &-0.024\\
.75 &0.158 &-0.020\\
0.8 &0.154 &-0.016\\
\hline\end{tabular}
\caption{ Local exchange splitting values (in Ryd) in
Cr$_{x}$Fe$_{1-x}$ alloys with varying
concentration of Cr}
\end{table}

The charge-redistribution procedure in this case is quite different. Like
FeCo, here the disorder in minority bands is stronger as is seen from the
$\delta^{\sigma}$ values. $\delta^{\downarrow}$ varies from 9 mRyd to
2 mRyd from Fe-rich to Cr-rich side while $\delta^{\uparrow}$ remains
around a value of -0.15 Ryd. This stronger disorder in minority bands
indicate a localization of majority electrons. As is seen from the density
of states  (Figure 11), both the bonding and antibonding part of spin down
density of states is dominated by Cr.In case of majority band the bonding
part is dominated By Fe and antibonding by Cr.
Along with this the nature of variation
of number of electrons for both spins at both the constituents  (Figure 10)
suggest that in this case, unlike the previous two, the electron redistribution
occurs mainly between Cr$^{\uparrow}$ and Cr$^{\downarrow}$ states. 
Electrons from Cr minority band migrate to Cr majority band explaining the
rapid increase of Cr moment.

Figure 12  (b) shows the variation of inverse spin susceptibility of the
system with increasing Cr content. Since n ($E_{F}$) has a maxima around
$x$=0.4,  this curve shows a minima around the same region. However, in the
case of spin fluctuation temperature variation (Figure 12  (c)) there is no
signature of this nature because the effect of magnetic moment variation is
much stronger across the concentration region and hence the variation of
Spin fluctuation temperature reflects the nature of variation of magnetic
moment only. As is seen in the previous two systems,  the MW Curie 
temperature too has a similar nature of variation (shown by solid line in Figure
12  (d)) while $T_{C}^{S}$ values are much higher.

\section{Conclusions}
We have studied the magnetism in bcc based Fe alloys where the three
constituents alloyed with Fe belong to the same row of the periodic table and consecutive
nearest neighbours of Fe. We have restricted ourselves to the ferromagnetic
regions of these alloys only. Our study reveals quite different natures
of electronic redistributions among the constituents as we go along from
Co to Cr producing different nature of variation of magnetization,  
Curie temperature and spin susceptibility. We have shown the dominant role of 
hybridization and magnetic exchange under the constraint of local charge
neutrality to explain successfully the variations in magnetic properties
of the alloys of nearest neighbours in periodic table.

\section*{Acknowledgements}
CBC would like to thank the CSIR, India for financial assistance.

\newpage
\section*{Figure Captions}
\begin{description}
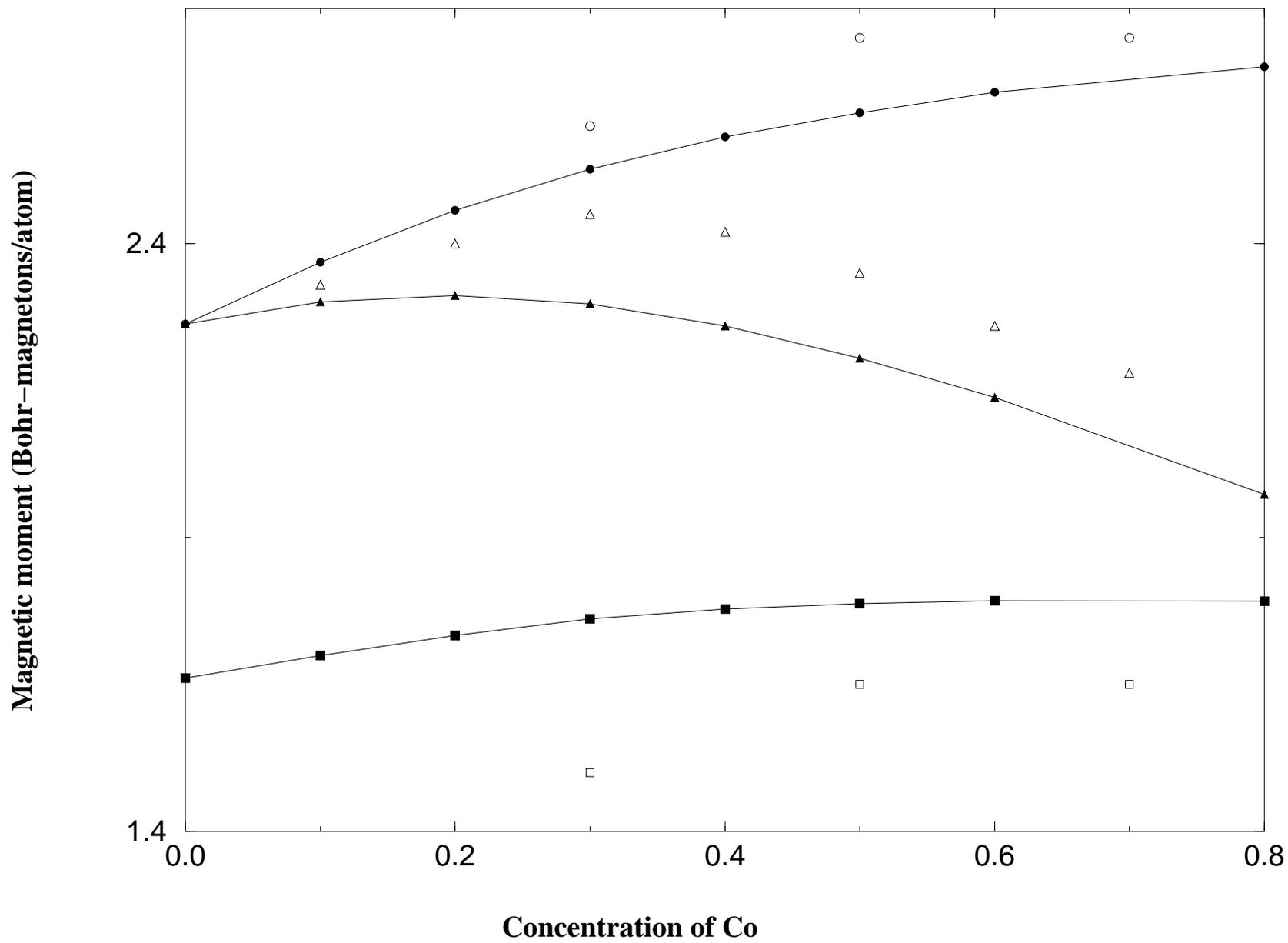
\item[Figure 1] Partial and averaged magnetic moments
(in Bohr-magnetons/atom)
vs. concentration of Co for Co$_{x}$Fe$_{1-x}$ alloy. The solid line with
 filled up triangles represents
 calculated averaged values, the solid line with filled
circles represents calculated Fe moments, the solid line with filled squares
represents calculated Co moments. The open up triangles are the experimental
values of average moments, the open circles are the experimental Fe moments and
the open squares are the experimental Co moments.
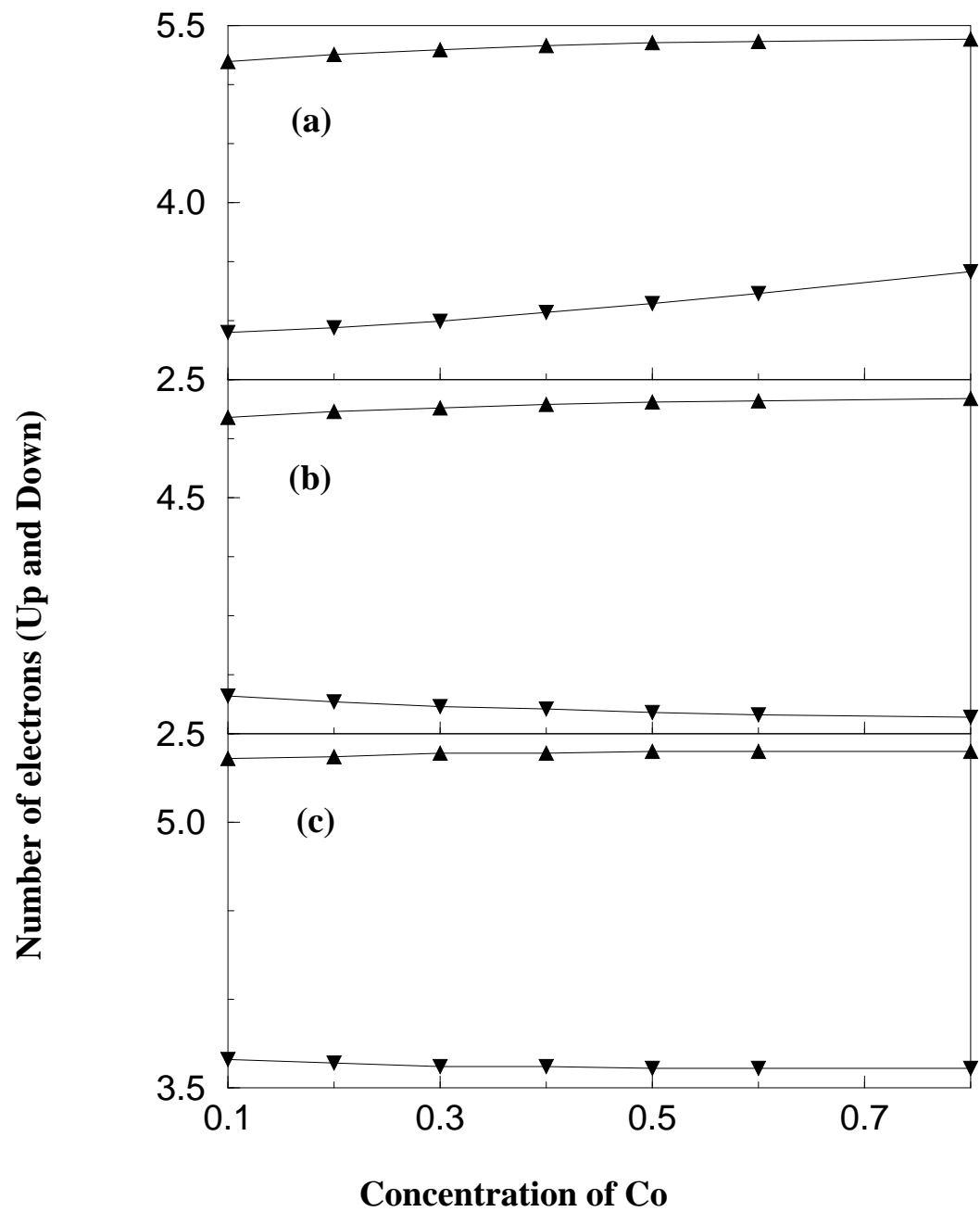
\item[Figure 2]   (a) represents average number of valence electrons for both
spins vs. concentration of Co for Co$_{x}$Fe$_{1-x}$ alloy. 
 (b) represents number of electrons at Fe site for both spins
vs. concentration of Co. (c) represents number of electrons at Co site
for both spins.
The up and the down triangles
are for the spin-up and for the spin-down electrons respectively for all the
three cases.
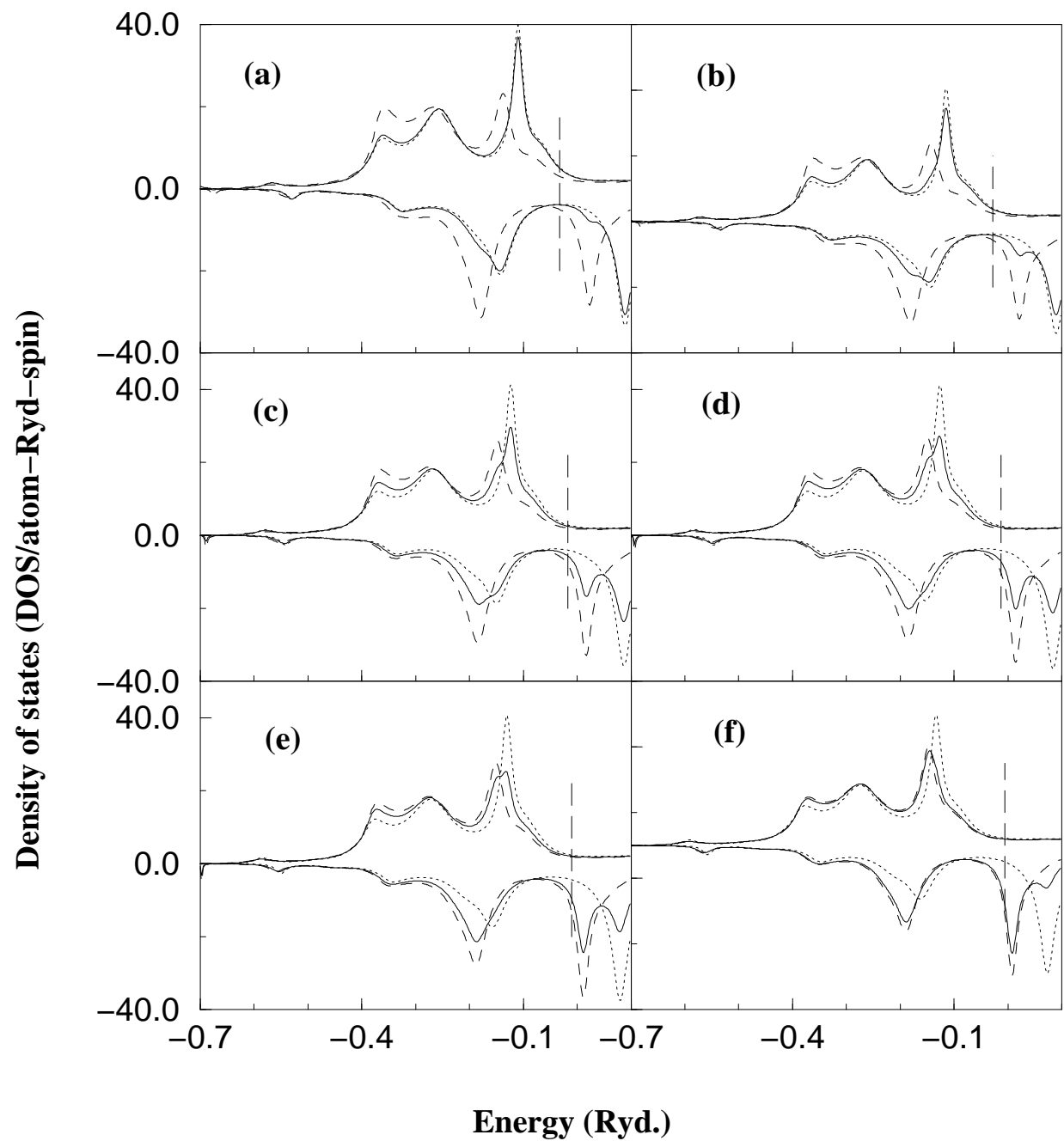
\item[Figure 3] Spin projected partial and averaged densities of states of
Co$_{x}$Fe$_{1-x}$ alloy. The various panels are for different concentrations
of Co-  (a)10$\%$   (b)20$\%$   (c)40$\%$   (d)50$\%$   (e)60$\%$   (f)80$\%$. 
In all the cases, 
the solid line represents averaged density of states while the dashed and
the dotted line stand for Co and Fe partial density of states respectively.
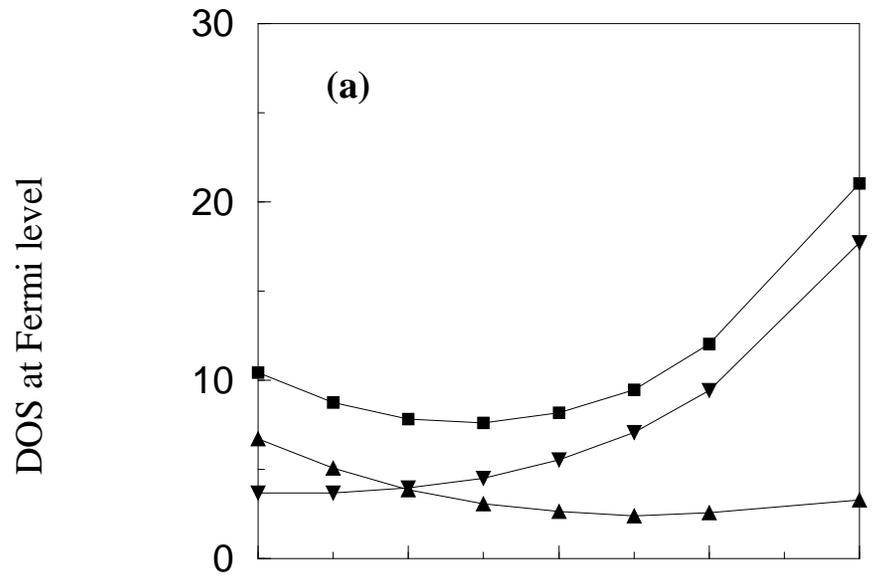
\item[Figure 4]   (a) represents density of states at Fermi level for both
spins as well as the averaged values vs. concentration of
Co in Co$_{x}$Fe$_{1-x}$ alloy. The up and down triangles stand
for values of up and down spins
respectively while the squares stand for averaged values.   (b) represents
averaged inverse spin susceptibilities
$ \left  ({2\mu_{B}^2}\right/{\chi}$ in
Ryd-atom vs. concentration of Co in Co$_{x}$Fe$_{1-x}$ alloy. The circles
represent the calculated values.   (c) represents
variation of spin fluctuation
temperature  ($T_{SF}$) in Kelvin vs.
concentration of Co in Co$_{x}$Fe$_{1-x}$ alloy.
  (d) the solid line and the filled circles represent the variation of Curie
temperature  ($T_{C}$) in Kelvin calculated using MW model
vs Co concentration. The diamonds
stand for the values of Stoner Curie temperature  ($T_{C}^{S}$) in Kelvin.
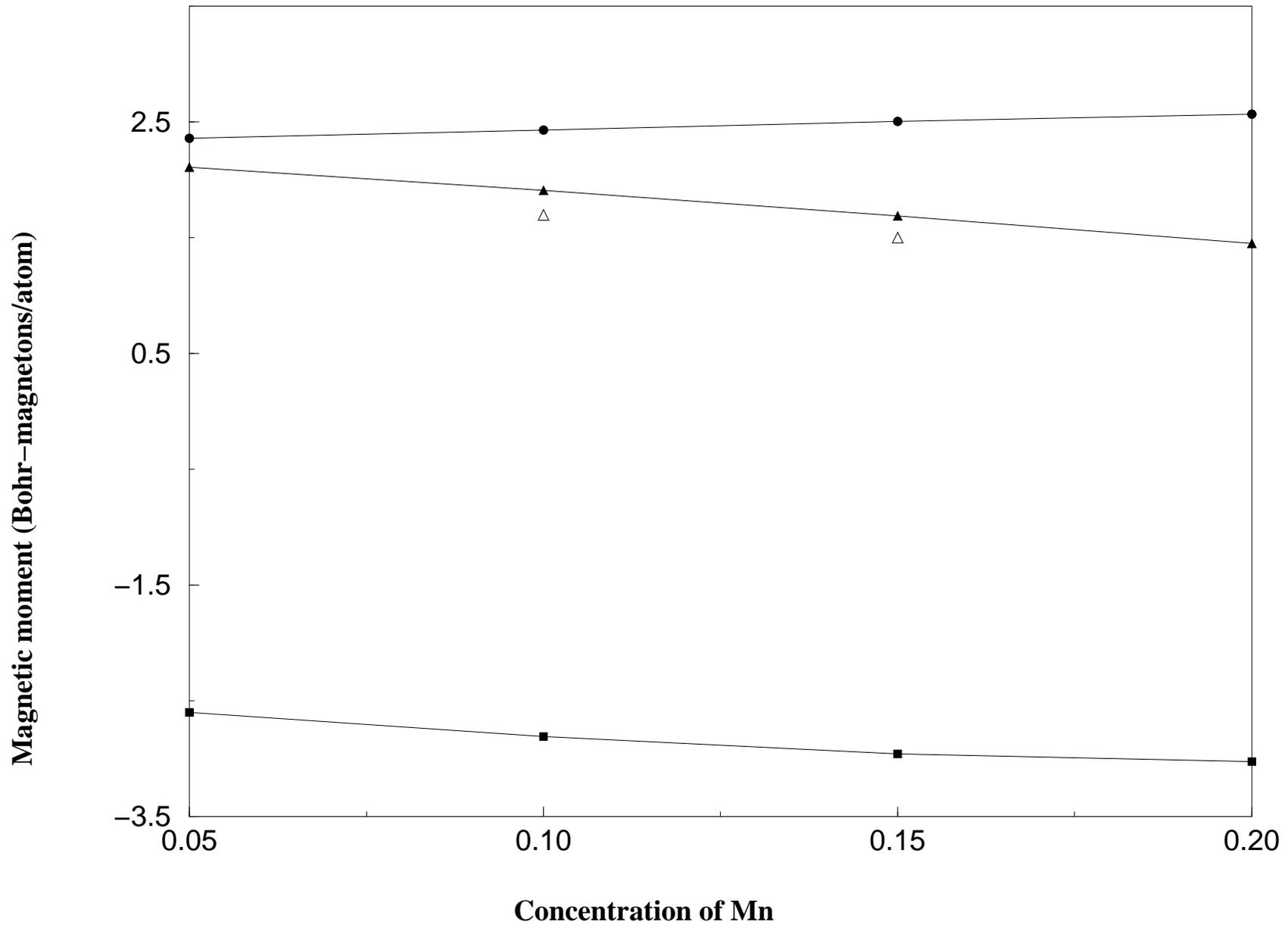
\item[Figure 5] Partial and averaged magnetic moments
(in Bohr-magnetons/atom)
vs. concentration of Mn for Mn$_{x}$Fe$_{1-x}$ alloy. The solid line with
up triangles represents calculated averaged values, the solid line with filled
circles represents calculated Fe local
moments, the solid line with filled squares
represents calculated Mn local moments.
The open up triangles are for the experimental
values of the average moments.
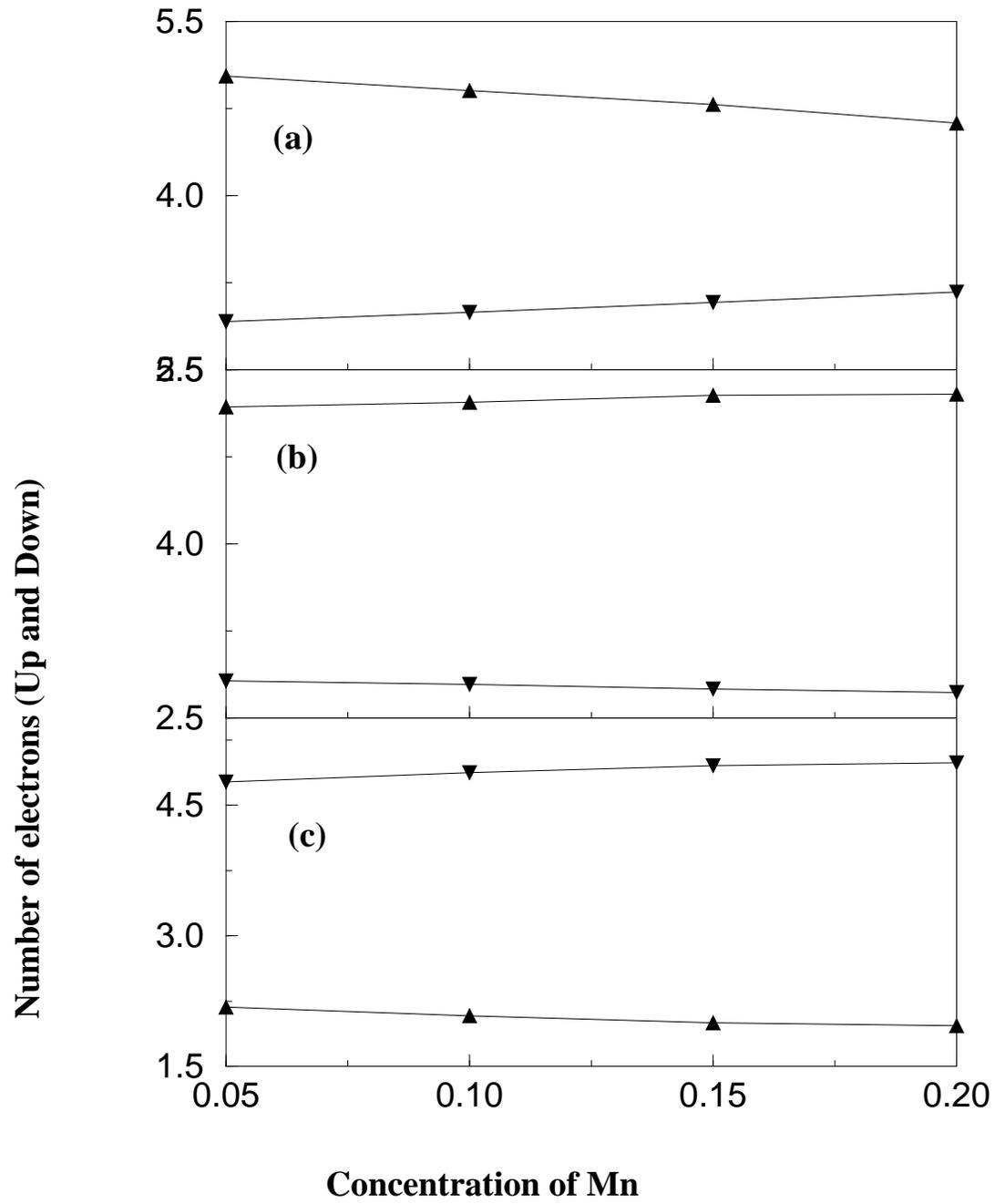
\item[Figure 6]   (a) represents average number of valence electrons for both
spins vs. concentration of Mn for Mn$_{x}$Fe$_{1-x}$ alloy. 
(b) represents number of electrons at Fe site for both spins
vs. concentration of Mn. 
(c) represents number of electrons at Mn site
for both spins. 
The up and the down triangles
are for the spin-up and for the spin-down electrons respectively for all the
three cases.

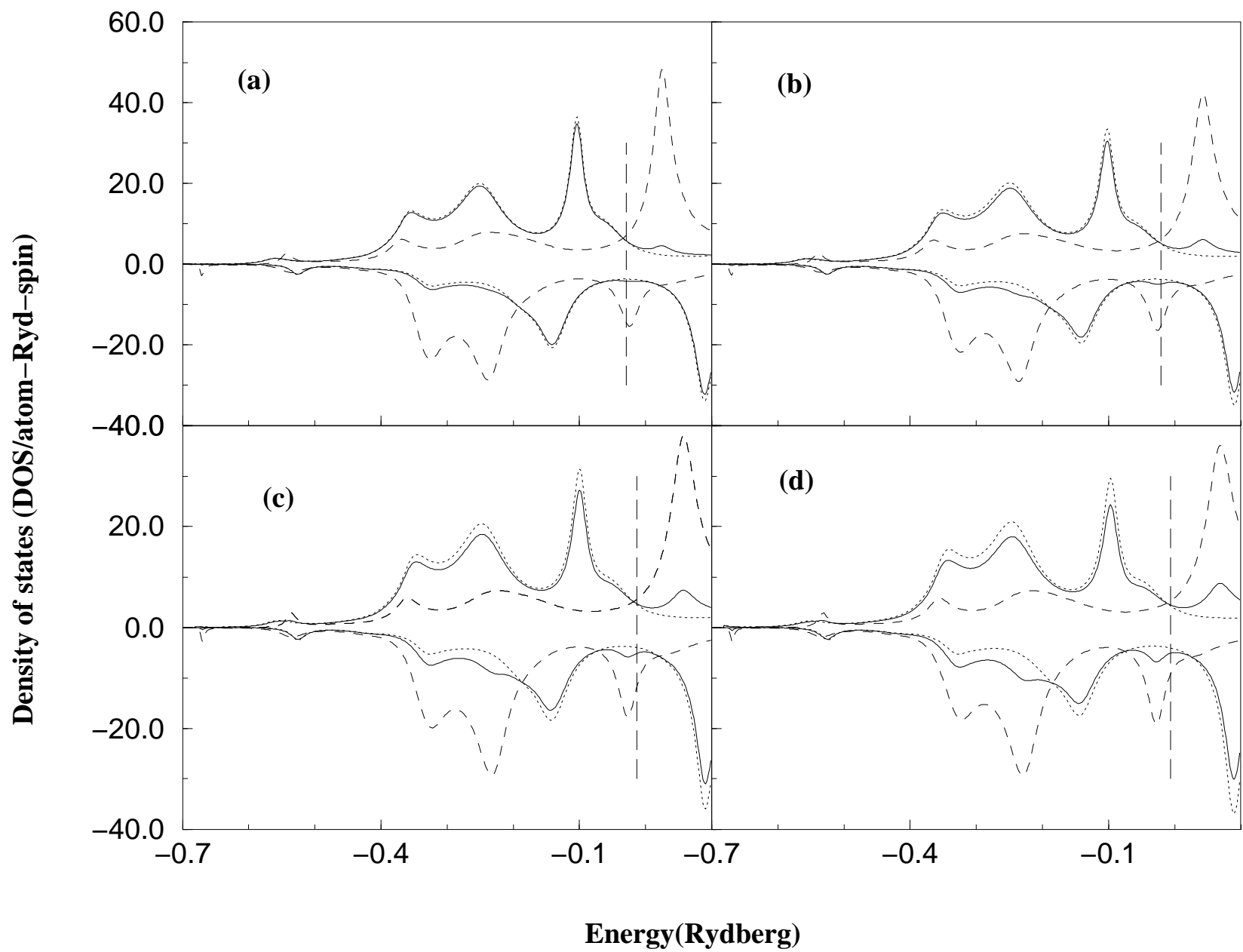
\item[Figure 7] Spin projected partial and averaged densities of states of
Mn$_{x}$Fe$_{1-x}$ alloy. The various panels are for different concentrations
of Mn-  (a)5$\%$   (b)10$\%$   (c)15$\%$   (d)20$\%$ . 
In all the cases, 
the solid line represents averaged density of states while the dashed and
the dotted line stand for Mn and Fe partial density of states respectively.
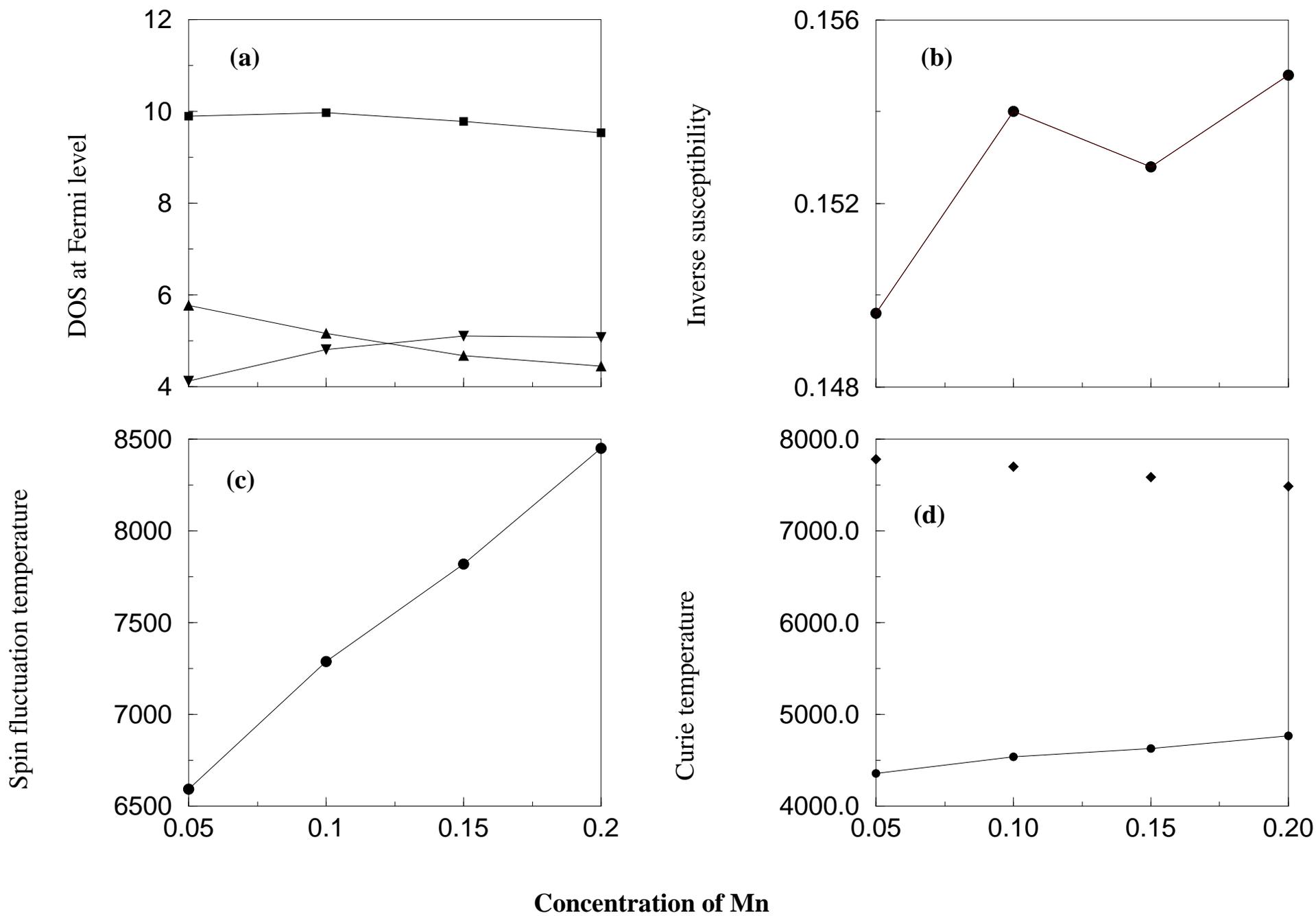
\item[Figure 8]   (a) represents density of states at Fermi level for both
spins as well as the averaged values vs. concentration of Mn in Mn$_{x}$Fe$_{1-x}$ alloy. The up and down triangles stand for values of up and down spins
respectively while the squares stand for averaged values.   (b) represents
averaged inverse spin susceptibilities $ \left  ({2\mu_{B}^2}\right/{\chi}$ in
Ryd-atom vs. concentration of Mn in Mn$_{x}$Fe$_{1-x}$ alloy. The circles
represent the calculated values.   (c) represents variation of spin fluctuation
temperature  ($T_{SF}$) in kelvin vs concentration of Mn in Mn$_{x}$Fe$_{1-x}$
  (d) the solid line and the filled circles represent the variation of Curie
temperature  ($T_{C}$) in Kelvin calculated using MW model vs.
 Mn concentration. The diamonds
stand for the values of Stoner Curie temperature  ($T_{C}^{S}$) in Kelvin.
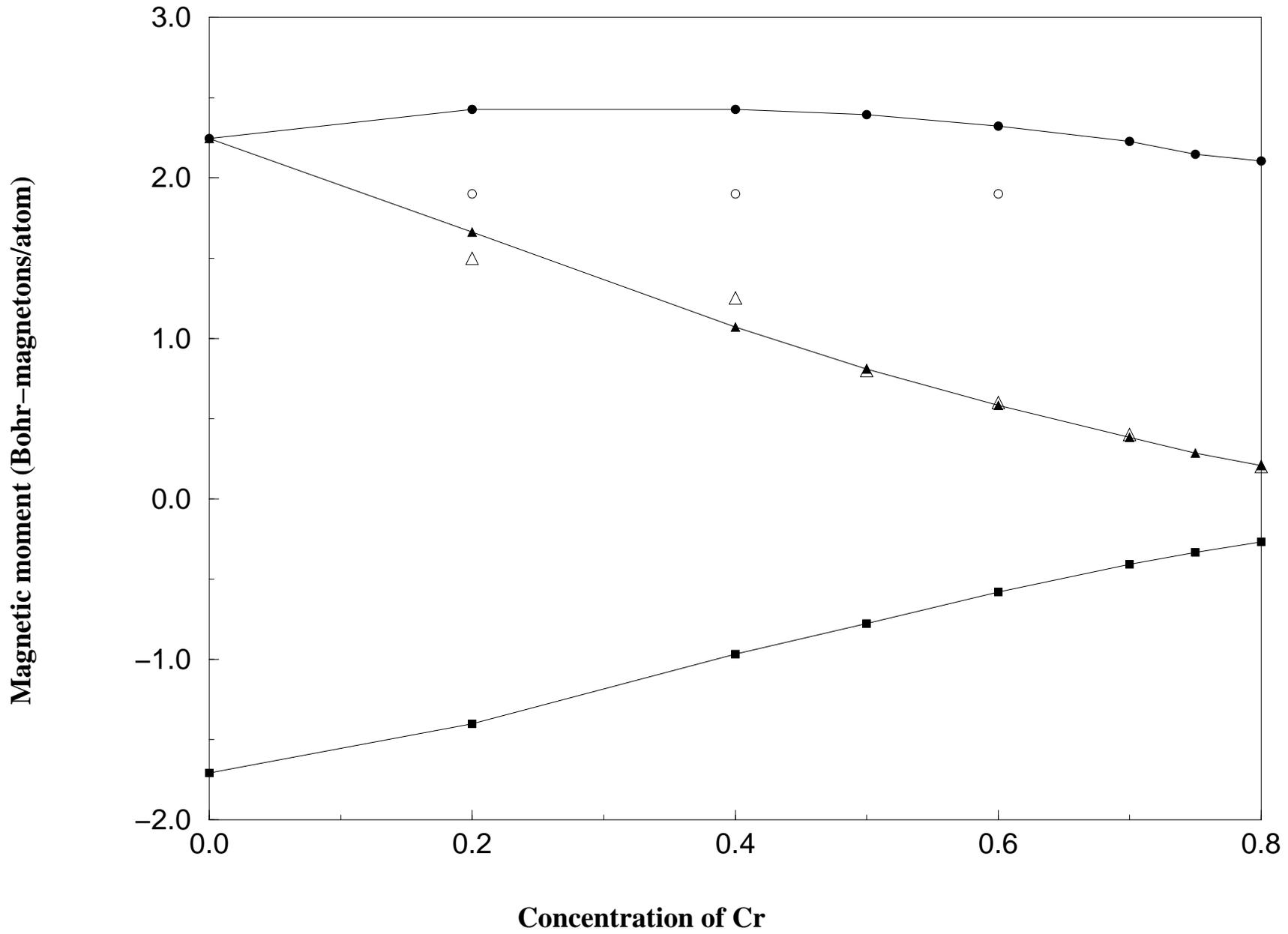
\item[Figure 9] Partial and averaged magnetic moments
(in Bohr-magnetons/atom)
vs. concentration of Cr for Cr$_{x}$Fe$_{1-x}$ alloy. The solid line with
up triangles represents calculated averaged values, the solid line with filled
circles represents calculated Fe local
moments, the solid line with filled squares
represents calculated Cr local moments. The open up triangles are for
the experimental
values of average moments and the open circles are for
the experimental Fe moments.
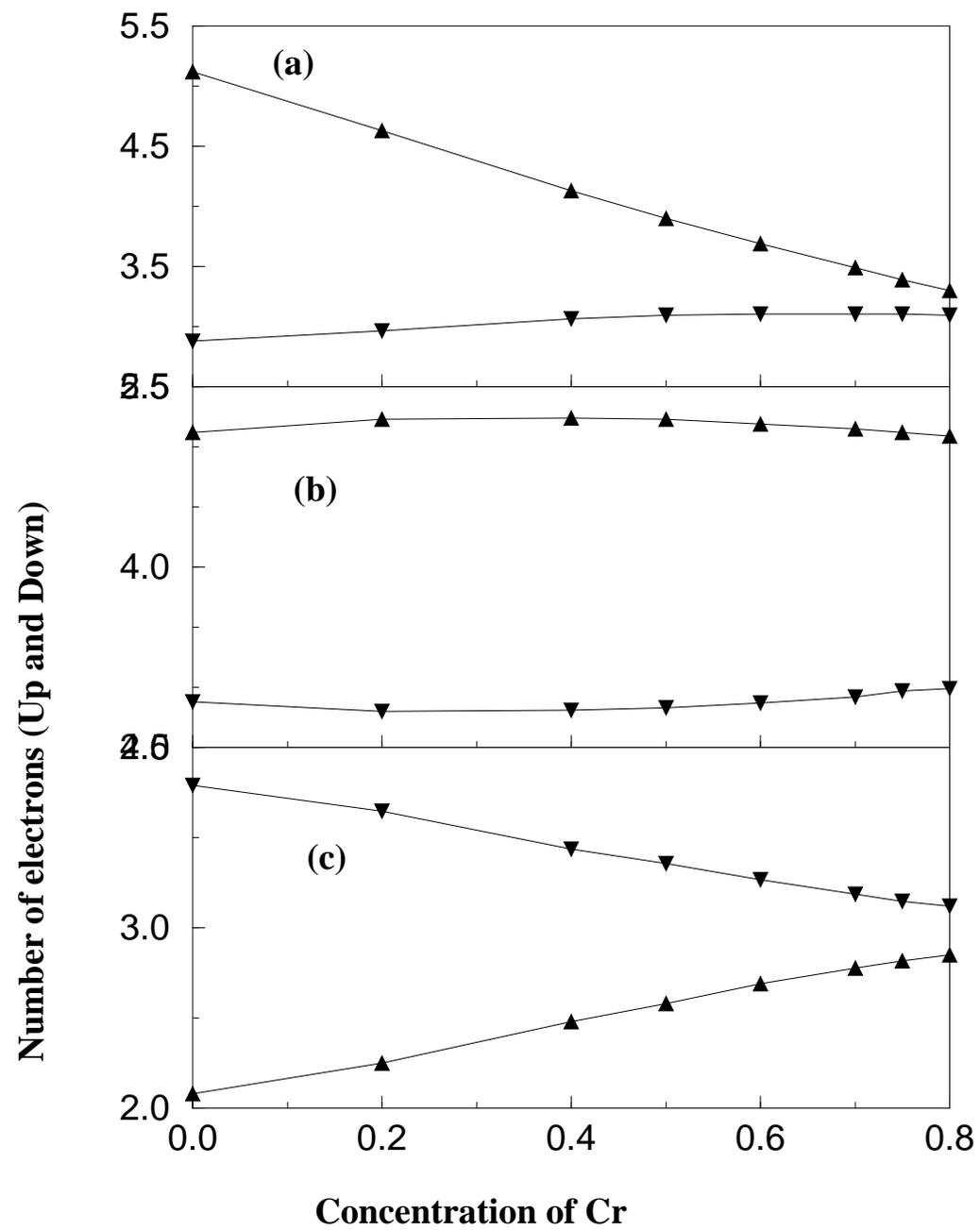
\item[Figure 10]   (a) represents average number of valence electrons for both
spins vs. concentration of Cr for Cr$_{x}$Fe$_{1-x}$ alloy. 
(b) represents number of electrons at Fe site for both spins
vs. concentration of Cr. 
(c) represents number of electrons at Cr site
for both spins.
The up and the down triangles
are for the spin-up and for the spin-down electrons respectively for all the
three cases.
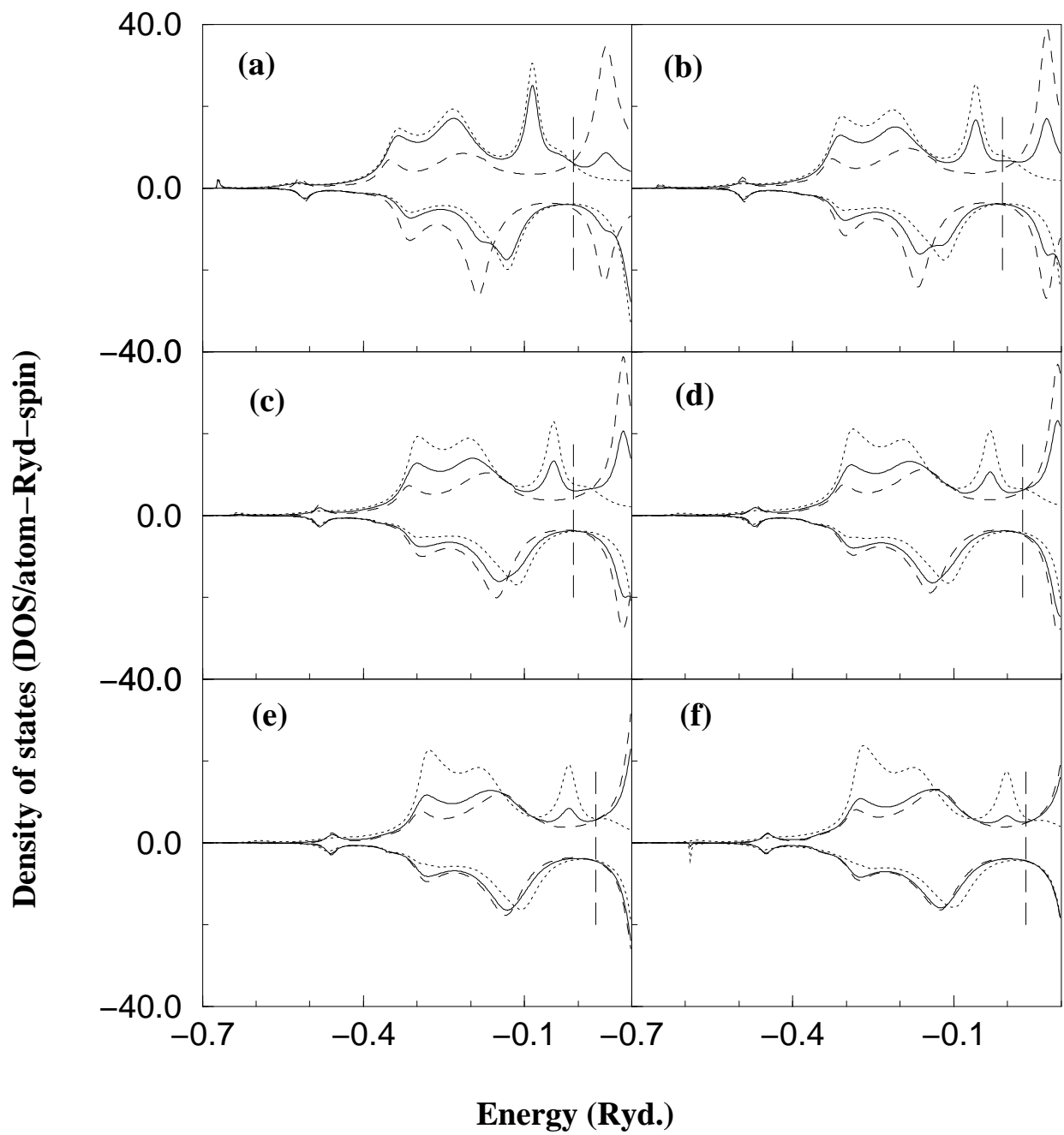
\item[Figure 11] Spin projected partial and averaged densities of states of
Cr$_{x}$Fe$_{1-x}$ alloy. The various panels are for different concentrations
of Cr-  (a)25$\%$   (b)40$\%$   (c)50$\%$   (d)60$\%$   (e)70$\%$   (f)80$\%$. 
In all the cases, 
the solid line represents averaged density of states while the dashed and
the dotted line stand for Cr and Fe partial density of states respectively.
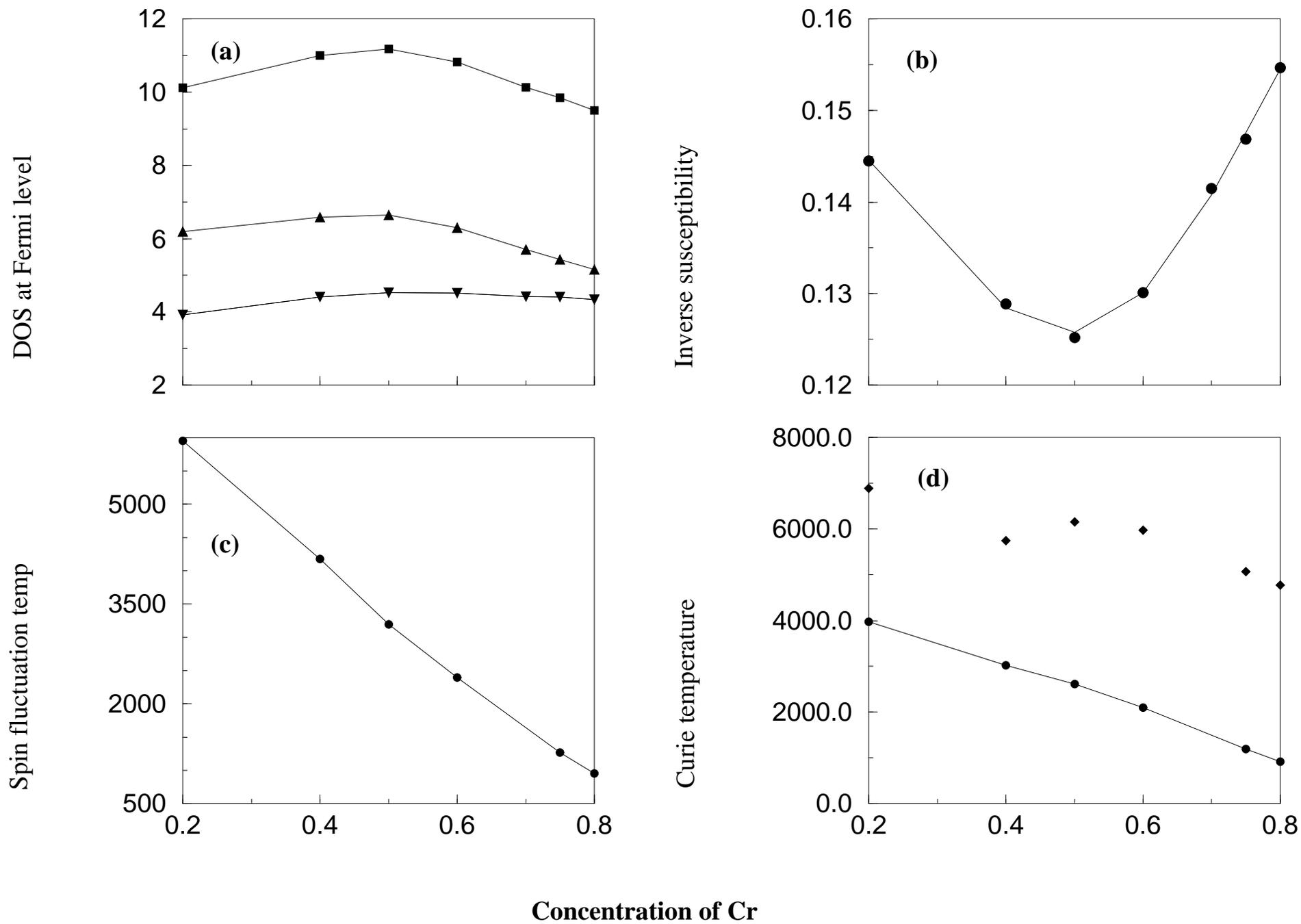
\item[Figure 12]   (a) represents density of states at Fermi level for both
spins as well as the averaged values vs. concentration of Cr in Cr$_{x}$Fe$_{1-x}$ alloy. The up and down triangles stand for values of up and down spins
respectively while the squares stand for averaged values.   (b) represents
averaged inverse spin susceptibilities $ \left  ({2\mu_{B}^2}\right/{\chi}$ in
Ryd-atom vs. concentration of Cr in Cr$_{x}$Fe$_{1-x}$ alloy. The circles
represent the calculated values.   (c) represents variation of spin fluctuation
temperature  ($T_{SF}$) in Kelvin vs. concentration of Cr in Cr$_{x}$Fe$_{1-x}$
  (d) the solid line and the filled circles represent the variation of Curie
temperature  ($T_{C}$) in Kelvin calculated using
MW model vs Cr concentration. The diamonds
stand for the values of Stoner Curie temperature  ($T_{C}^{S}$) in Kelvin.

\end{description}


\section*{References}
\begin{thebibliography}{99}
\bibitem{kn:wohl} Ferromagnetic Materials edited by E.P. Wohlfarth and K.H.J.Buschow  (North-Holland, Amsterdam, 1980-1993), Vol. 1-7
\bibitem{kn:ishi} Ishikawa Y. in Physics and Applications of INVAR Alloys, 
{\it edited} by H. Saito, Honda Memorial Series on Materials Science No. 3
  (Moruzen, Tokyo, 1978); Hasegawa H. and Pettifor D.G. 1983 \PRL {\bf 50} 130
\bibitem{kn:and} Andersen O.K. {\it et al} 1977 Physica {\bf B+C 86-88}
249-256; Kubler J. 1981 {\it Physics Letters} {\bf 81A} 81; Wang C.S., 
Klein B.M. and Krakauer H. 1985 \PRL {\bf 54} 1852
\bibitem{kn:pinski} Pinski F.J. {\it et al} 1986 \PRL {\bf 56} 2096; Gonser
U., Krische K. and Nasu S. 1980 {\it Journal of Magnetism and Magnetic Materials}
{\bf 15-18} 1145; Tsunoda Y. 1988 \JPCM {\bf 1} 10427; Moruzzi V.L.
{\it et al} 1986 \PR {\bf B34} 1784
\bibitem{kn:hirai} Hirai K. 1989 {\it Journal of Phys. Soc. Japan} {\bf 58}
 4288; Mryasov O.N. {\it et al} 1991 \JPCM {\bf 3} 7683; Lichtenstein A.I., 
Katznelson M.I. and Gubanov V.A. 1984 \JPF {\bf 14} L125; Mryasov O.N., 
Gubanov V.A. and Lichtenstein A.I. 1992 \PR {\bf B45} 12330
\bibitem{kn:shull} Shull C.G. and Wilkinson M.K. 1955 \PR {\bf 97} 304
\bibitem{kn:bardos} Bardos D.I. 1969 {\it J. Appl. Phys.} {\bf 40} 1371
\bibitem{kn:collins} Collins M.F. and Low G.G. 1965 {\it Proc. Phys. Soc.}
 {\bf 86} 535
\bibitem{kn:child} Child H.R. and Cable J.W. 1976 \PR {\bf B13} 227
\bibitem{kn:radha} Radhakrishnan P. and Livet F. 1978 {\it Solid state comm.}
{\bf 25} 597
\bibitem{kn:kaj} Kajzar F. and Parette G. 1980 \PR {\bf B22} 5471
\bibitem{kn:arp} Arp V., Edmomnds D. and Petersen R. 1952 \PRL {\bf 3} 212, 
Johnson C.E., Ridout M.S. and Cranshaw T.E. 1963 {\it Proc. Phys. Soc.} {\bf 81}
1079; Lutgemeier H. and Dubiel S.M. 1982 {\it Journal of Magnetism and Magnetic Materials} {\bf 28} 277; Muraoka Y. {\it et al} 1976 {\it J. Phys. Soc. Japan} {\bf 40} 414; Koi Y., Tsujimura T. and Hihara T. 1964 {\it J. Phys. Soc. Japan} {\bf 19} 1493
\bibitem{kn:stoel} Stoelinga J.H.M. and Gersdorf R. 1966 {\it Physics Letters}
{\bf 19} 640
\bibitem{kn:cheng} Cheng C.H., Wei C.T. and Beck P.A. 1960 \PR {\bf 2} 426
\bibitem{kn:schroder} Scroder K. 1962 \PR {\bf 125} 1209
\bibitem{kn:baal} Van Baal C.M. 1973 {\it Physica} {\bf 64} 571; Kikuchi R. and Sato H. 1974 {\it Acta Metall.} {\bf 22} 1099; Umebayashi H. and Ishikawa Y. 1966 {\it J. Phys. Soc. Japan} {\bf 21} 1281
\bibitem{kn:has} Hasegawa H. and Kanamori J. 1972 {\it J. Phys. Soc. Japan} {\bf 33} 1607; Mattews J.C. 1972 {\it J. Phys. Soc. Japan} {\bf 32} 110
\bibitem{kn:jo} Jo T. 1982 {\it J. Phys. Soc. Japan} {\bf 3} 794
\bibitem{kn:james} James P. {\it et al} 1999 \PR {\bf B59} 419; Schwarz K. {\it et al} 1984 \JPF {\bf 14} 2659
\bibitem{kn:burke} Burke S.K. and Rainford B.D. 1983 \JPF {\bf 13} 441; Ebert H. {\it et al} 1990 \JPCM {\bf 2} 443; Martinez-Herrera F.J. {\it et al} 1985 \PR {\bf B31} 1686; Drittler B. 1989 {\it et al} \PR {\bf B40} 8203
\bibitem{kn:akai} Akai H. and Dederichs P.H. 1993 \PR {\bf B47} 8739; Bluegel S. {\it et al} 1987 \PR {\bf B35} 3271; Turek I. {\it et al} 1994 \PR {\bf B49} 3352
\bibitem{kn:pad} Paduani C. and Krause J.C. 1998 \PR {\bf B58} 175; Dederichs P.H. {\it et al} 1991 {\it Journal of Magnetism and Magnetic Materials} {\bf 100} 241; Kaspar J. and Salahub D.R. 1983 \JPF {\bf 13} 311; Qiu S.L., Marcus P.M. and Moruzzi V.L. 1998 \PR {\bf B58} 2651
\bibitem{kn:kuli} Kulikov N.I. and Tugushev V.V. 1984 {\it Sov. Phys. Usp} {\bf 27} 954; Fawcett E. {\it et al} 1994 {\it Rev. Mod. Phys.} {\bf 66} 25; Furuska M. {\it et al} 1986 {\it J. Phys. Soc. Japan} {\bf 55} 2253
\bibitem{kn:endoh} Endoh Y. and Ishikawa Y. 1971 {\it J. Phys. Soc. Japan} {\bf 30} 1614
\bibitem{kn:as} Mookerjee A. and Prasad R. 1993 \PR {\bf B48} 17724 
\bibitem{kn:saha} Saha T., Dasgupta I. and Mookerjee A. 1994 \PR {\bf B50} 13267
\bibitem{kn:bs} Sanyal B. {\it et al 1998} \JPCM {\bf 10} 5767
\bibitem{kn:oka} Andersen O K,  Jepsen O and Glotzel 1985 {\it Highlights of Condensed-Matter Theory},  edited by Bassani F,  Fumi F
and Tosi M P   (North-Holland,  New York),  p. 59
\bibitem{kn:ppb} Biswas P.P. {\it et al} 1995 \JPCM {\bf 7} 8569
\bibitem{kn:ln} Luchini M U and Nex C M M 1987 \JPC {\bf 20} 3125
\bibitem{kn:gdm} Ghosh S,  Das N and Mookerjee A 1997 \JPCM {\bf 9} 1701
\bibitem{kn:mw} Mohn P H and Wolfarth E P 1987 {\it J. Phys. F} {\bf 17} 2421
\bibitem{kn:jan} Janak J.F. 1977 \PR {\bf B16} 255
\bibitem{kn:gun} Gunnarson O 1976 {\it J. Phys. F} {\bf 6} 587
\bibitem{kn:ger} Gersdorf R 1962 {\it J. Phys. Radium} {\bf 23} 726
\bibitem{kn:sdm} Saha T,  Dasgupta I and Mookerjee A 1996 \JPCM {\bf 8} 1979
\bibitem{kn:gm} Ghosh S., Das N. and Mookerjee A. 1999 {\it Modern Phys. Letters} {\bf B13} 723
\bibitem{kn:kur} Turek I. {\it et al} in {\it Electronic structure of
disordered alloys, surfaces and interfaces} 1997, Kluwer Academic Publishers
\bibitem{kn:rich} Richter R. and Eschrig H. 1988 \JPF {\bf 18} 1813
\bibitem{kn:schwarz} Schwarz K. and Salahub D.R. 1982 \PR {\bf B25} 3427
\bibitem{kn:coll} Collins M.F. and Forsyth J.B. 1963 {\it Phil. Mag.} {\bf 8} 401
\bibitem{kn:turek} Turek I. {\it et al} 1994 \PR {\bf B49} 3352
\bibitem{kn:rich1} Richter R. and Eschrig H. 1987 {\it Proc. 7th General EPS
conf.}, Pisa
\bibitem{kn:fish} Fisher H. {\it et al} in {\it Magnetic Ultrathin Films, Multilayers and Surfaces},  edited by A. Fert {\it et al},  MRS symposia proceedings No. 384  (Materials Research Society, Pittsburgh, 1995)
\bibitem{kn:kan1} Hasegawa H. and Kanamori J. 1972 {\it J. Phys. Soc. Japan}
{\bf 33} 1607
\bibitem{kn:kuli1} Kulikov N.I. and Demangeat C. 1997 \PR {\bf B55} 3533
\bibitem{kn:ling} Ling M.F., Staunton J.B. and Johnson D.D. 1995 \JPC {\bf 7}
1863
\bibitem{kn:but} Buttler W.H. {\it et al} 1995 {\it Journal of Magnetism and
Magnetic Materials} {\bf 151} 354
\bibitem{kn:ded} Dederichs P.H. {\it et al} 1991 {\it Journal of Magnetism
and Magnetic Materials} {\bf 100} 241

\end{thebibliography}
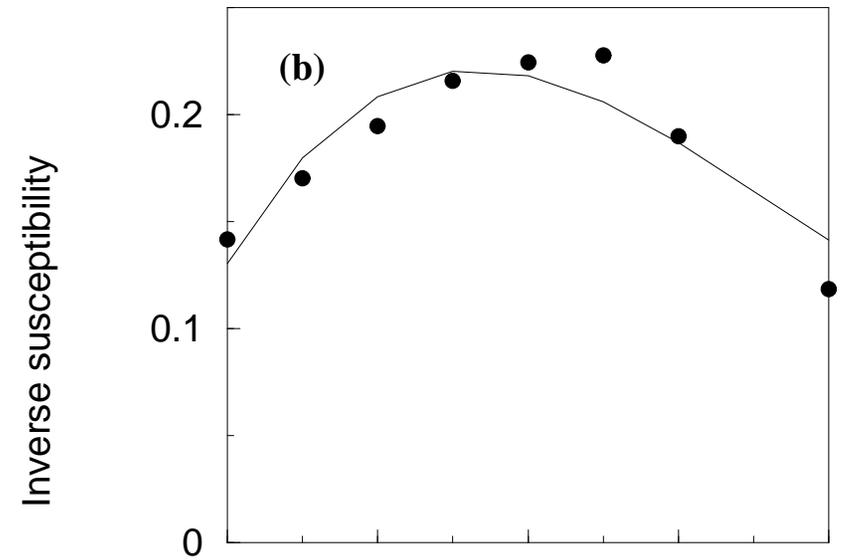
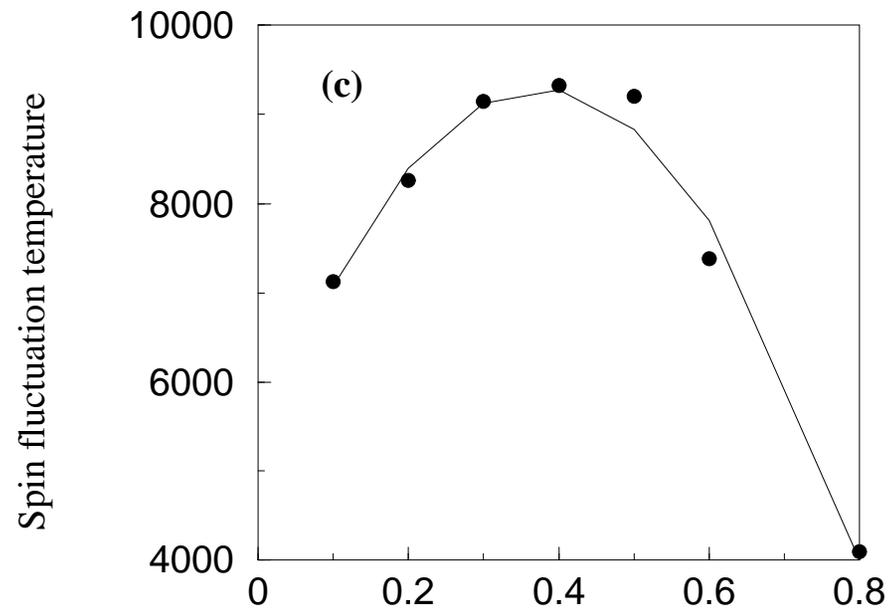
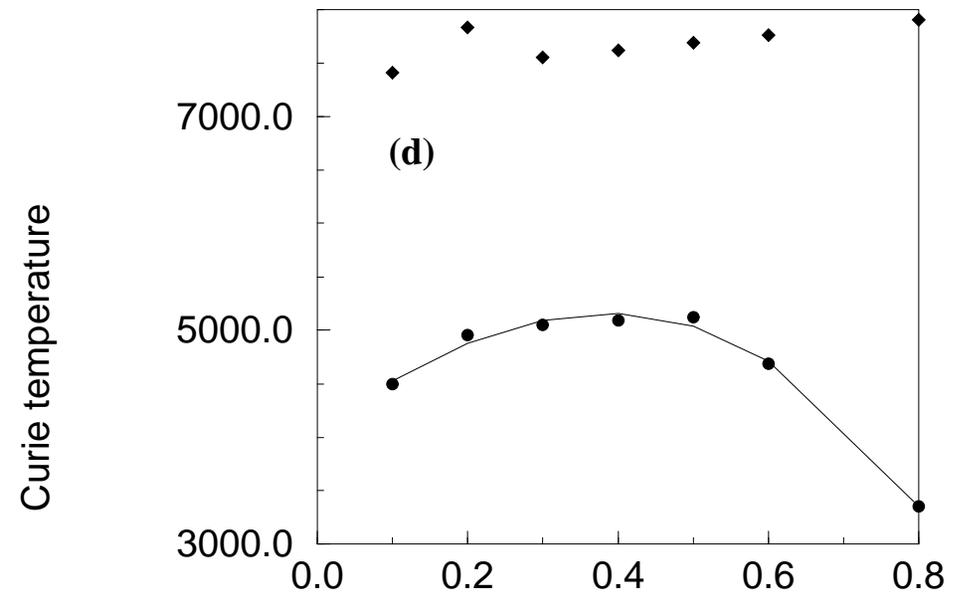
\end{document}